\documentclass{PoS}

\title{STEREO: Search for sterile neutrinos at the ILL}

\ShortTitle{STEREO: Search for sterile neutrinos at the ILL}

\author{\speaker{Luis Manzanillas}  \thanks{On behalf of the STEREO collaboration}\\
        LAPP - IN2P3 - CNRS / Universit\'e Savoie Mont Blanc - Universit\'e Grenoble Alpes\\
        E-mail: \email{lmanzanillas@gmail.com}}

\abstract{Recent studies have shown that there are discrepancies between 
observations and the theoretical predictions in some neutrino  experiments at short
distances. In the so-called ``Reactor Antineutrino Anomaly'' and in the ``Gallium Anomaly'', these 
differences from the expectations are at the $\sim$3 $\sigma$ level in both cases. Oscillations
into a light sterile neutrino state ($\Delta m^{2} \sim 1eV^{2}$) could account for the deficits in 
observed rates. The STEREO experiment has been conceived to 
confirm or reject the sterile neutrino hypothesis. It will search 
for an oscillation pattern at short baselines (9-11 m) in the energy spectrum of the
antineutrinos emitted by the research nuclear reactor of the Institut Laue Langevin. To 
this end, the detector is filled with two tons of Gd-loaded liquid scintillator read 
out by an array of PMTs and is segmented into 6 cells in the direction of the antineutrino's 
propagation. STEREO should be capable of excluding the best fit parameters region of a candidate sterile neutrino at 5 $\sigma$. Data 
taking will start at the end of 2016 and the first physics results are expected by the first semester of 2017.}

\FullConference{Neutrino Oscillation Workshop\\
		4 - 11 September, 2016\\
		Otranto (Lecce, Italy)}

\begin{document}

\section{Introduction}
During the preparation of the current $\nu$ experiments devoted to the measurement of $\theta_{13}$ (Double Chooz,
Daya Bay and RENO), the $\nu$ spectra and fluxes of nuclear reactors were reevaluated. The new estimations
increased the predicted antineutrino flux by about 3 \% \cite{Mueller:2011nm,Huber:2011wv}.
Using the new estimations, a reevaluation of previous experiments' observed antineutrino flux, a total deficit
of around 7\% (3$\sigma$) was found \cite{Mention:2011rk}.
This is known as the reactor antineutrino anomaly (RAA).
Oscillations into a light sterile $\nu$ could account for 
such deficits \cite{Abazajian:2012ys}.
In addition, other anomalies pointing to the sterile $\nu$
hypothesis have also been observed in other experiments, which are known as the Gallium and
LSND anomalies \cite{Giunti:2006bj}. 

In the 3+1 $\nu$ framework, the oscillation probability induced by a sterile $\nu$ state at the eV$^2$ scale
can be reduced to the two flavor approximation. Thus, the survival probability is given by \cite{Giunti:2007ry}
\begin{equation}\label{ProbaOsc}
 P^{3+1}_{\overline{\nu_{e}}\rightarrow\overline{\nu_{e}}} 
 \simeq 1-\sin^{2}2\theta_{ee}\sin^{2} \left(\frac{\Delta m^{2}_{41}L}{4E_{\nu}}\right)
\end{equation} 
where $\theta$ and $\Delta m^{2}$ are respectively the mixing angle between the two flavors and
the difference of squared masses $m^{2}_{1}-m^{2}_{4}$, both being parameters given by Nature. $E_{\nu}$ is 
the $\nu$ energy and $L$ is the distance traveled by the $\nu$  from the source to the detection point.
In most of the reactor experiments, the $\overline{\nu_{e}}$ energy spectrum is measured at a 
fixed distance. However, since the uncertainties in the prediction of this spectrum 
are high, searching for a new oscillation requires an unambiguous 
signature: the distortion of this energy spectrum as function of the distance.
The best fit values of $\theta$ and $\Delta m^{2}$ suggested by the reactor and gallium 
anomalies \cite{Mention:2011rk} ($\Delta m^{2}\simeq 2.3$  $eV^{2}$, $\sin^{2}2\theta\simeq0.14$) implies an oscillation 
of about 3 m for $\overline{\nu_{e}}$'s of 3 MeV, which is a typical energy of detected reactor $\overline{\nu_{e}}$'s.
Thus, the other key ingredient to search for such oscillations is to have a compact source.

A broad experimental program is ongoing to test the sterile $\nu$ hypothesis using different detection techniques
and  $\nu$ sources. Among the experiments using a nuclear reactor as $\nu$ source we have 
the STEREO experiment, whose goal is to confirm or reject the existence
of a light sterile $\nu$ state, by searching for an oscillation pattern at short baselines in the
reactor $\overline{\nu_{e}}$ energy spectrum.
The source of $\overline{\nu_{e}}$'s for the STEREO experiment is
the most compact (37 cm diameter) research nuclear reactor in France, located at the 
Laue-Langevin Institute (ILL) in Grenoble. This reactor uses 
highly enriched $^{235}$U, which reduces the contribution from other fissile isotopes whose  $\overline{\nu_{e}}$
spectrum is less well known than $^{235}$U's. The center of STEREO is placed 10 m away from the reactor core. 

\section{The STEREO experiment}
\begin{figure}[h]
\begin{tabular}{cc}
\includegraphics[scale=0.25]{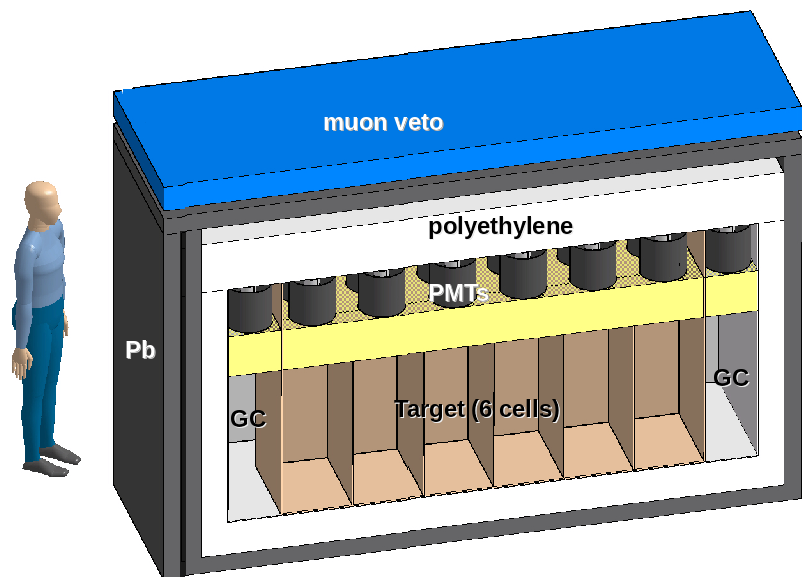}
&
\includegraphics[scale=0.04]{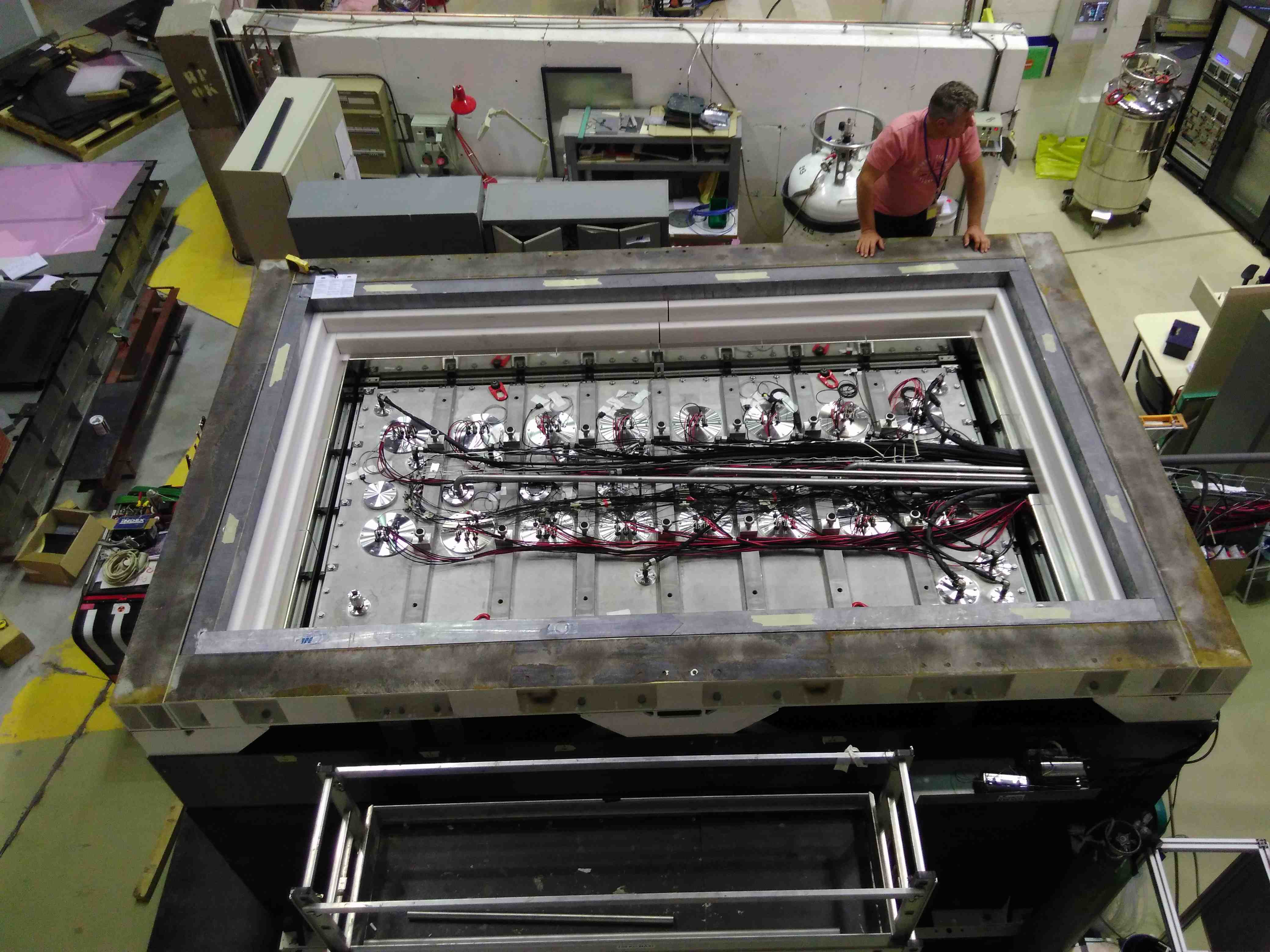}
\end{tabular}
\caption{Left: Sketch of the different parts of the STEREO 
experiment, which consists of a target of 6 cells filled with
LS doped with gadolinium, surrounded by an outer crown 
of LS called gamma catcher (GC). The scintillation light is detected by PMTs on top of the target cells
and the GC. Shielding 
is composed of polyethylene and lead. A muon detector is placed on the top. 
Right: Installation of STEREO at the ILL}
\label{fig2}
\end{figure}

STEREO will exploit a proven  $\overline{\nu_{e}}$ detection technology, using 2 tons of  Gd-loaded
liquid scintillator (LS). The detector is segmented into 
6 cells (40x90x90 cm$^{3}$) in the direction of the $\overline{\nu_{e}}$'s propagation 
(see figure \ref{fig2}). Each cell is read out by an array of 4 PMTs on the top. The target
cells are surrounded by an outer crown of 30 cm thickness called gamma catcher (GC), which 
is filled with LS without Gd. The main function of the GC is to improve the
energy resolution by recovering  some of the escaping gammas from the target cells. At the same 
time, the GC can be used as veto to remove backgrounds coming from the outside. 
In order to further protect the detector from fast and thermal neutrons and gammas, a passive shielding 
of polyethylene, lead and boron loaded rubber encloses the entire detector. Finally, to 
remove background events induced by cosmic muons, a water Cerenkov detector called ``muon veto'' (MV) is placed on top
of the detector. 

The $\overline{\nu_{e}}$'s are detected via the inverse beta decay (IBD) process:
$\overline{\nu_{e}}+p\rightarrow e^{+}+n$, which produces two signals correlated 
in time and distance. The positron deposits almost instantaneously its energy in
the LS, which, together with the deposited energy of the two 511 keV 
gammas from its annihilation with an electron, gives a prompt signal.
The neutron thermalizes and then diffuses until it's captured by a nucleus of Gd some $\sim$ 15
$\mu$s later, giving rise to a delayed signal, which consists of a gamma cascade
with a total energy of 8 MeV produced by the de-excitation of the Gd.

\section{Status of the experiment and expected physics results}
A dedicated electronics hosted in a $\mu$TCA crate was developed and tested using the MV
in 2015 \cite{Bourrion:2015axa}. 
In Spring of 2016 the inner detector (Target and GC) was assembled. During the 
summer the supporting structure and all the large items required for the STEREO shielding were delivered 
and mounted at the ILL. In August the inner detector was inserted inside the shielding structure and the calibration system installed.
Finally, the shielding roof was closed and the MV was 
installed at the beginning of September. The detector will be moved to the STEREO
site by the end of September and will be filled with LS at the beginning of November. Commissioning and  $\nu$
data taking are expected to start around the beginning of November with the new reactor cycle.

\begin{figure}[h]
\begin{center}
\includegraphics[scale=0.3]{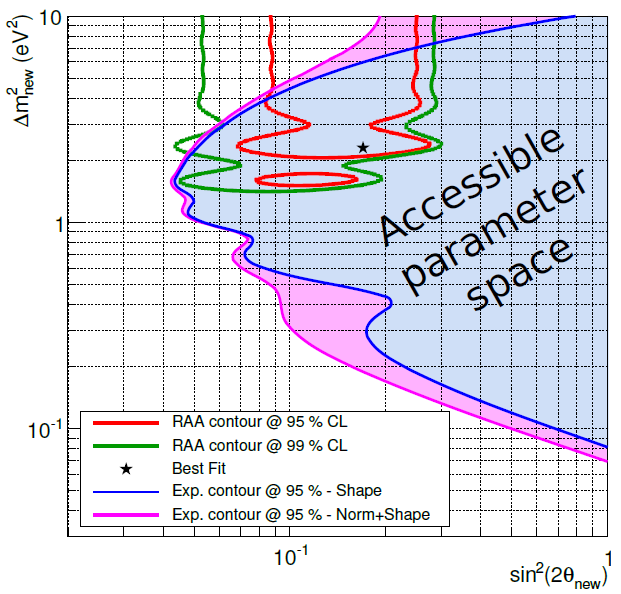}\end{center}
\caption{Expected exclusion contour of STEREO for 300 days of data taking}
\label{ContourStereo}
\end{figure}

The expected exclusion contour of STEREO for 300 days of data taking is shown in figure \ref{ContourStereo}. 
The main assumptions are: a signal to background ratio of 1.5, an uncertainty in the energy scale lower than 2 \%, an energy 
resolution of 10 \% at 2 MeV, and an energy threshold for the prompt and delayed signals of 2 and 5 MeV respectively. 
Most of the region of the parameter space allowed by the RAA  will be covered by STEREO. In addition, with around 400 $\nu$ events 
per day, STEREO will provide a new spectrum of reference for $^{235}$U with more than 100000 $\nu$'s. This new spectrum
could provide information to understand the origin of the bump around 5 MeV observed in the reactor 
$\overline{\nu_{e}}$ spectrum in current reactor $\nu$ experiments \cite{Buck:2015clx}. 
Before the spring shutdown of 2017, around 100 days of data taking are expected, enough to cover 
the best fit region of the RAA.

\section{Conclusions}
Most of the large items of STEREO have been delivered and installed at the ILL. The 
installation will be completed by October of 2016. The data taking will start by the 
beginning of November. The first physics results are expected by the first half of  2017
with around 100 days of data that can be accumulated before the spring shutdown 
of 2017. Thus, STEREO will provide the first search for sterile $\nu$'s with an
analysis of the reactor energy spectrum as function of the distance.


\begin{thebibliography}{99}

\bibitem{Mueller:2011nm} 
  T.~A.~Mueller {\it et al.},
  \emph{Improved Predictions of Reactor Antineutrino Spectra},
  Phys.\ Rev.\ C {\bf 83}, 054615 (2011)
  [arXiv:1101.2663].  
\bibitem{Huber:2011wv} 
  P.~Huber,
  \emph{On the determination of anti-neutrino spectra from nuclear reactors},
  Phys.\ Rev.\ C {\bf 84}, 024617 (2011)
  [arXiv:1106.0687].
\bibitem{Mention:2011rk} 
  G.~Mention, M.~Fechner, T.~Lasserre, T.~A.~Mueller, D.~Lhuillier, M.~Cribier and A.~Letourneau,
  \emph{The Reactor Antineutrino Anomaly},
  Phys.\ Rev.\ D {\bf 83}, 073006 (2011)
  [arXiv:1101.2755].
\bibitem{Abazajian:2012ys} 
  K.~N.~Abazajian {\it et al.},
  \emph{Light Sterile Neutrinos: A White Paper},
  [arXiv:1204.5379].
  
\bibitem{Giunti:2006bj} 
  C.~Giunti and M.~Laveder,
  \emph{Short-Baseline Active-Sterile Neutrino Oscillations?},
  Mod.\ Phys.\ Lett.\ A {\bf 22}, 2499 (2007)
  [hep-ph/0610352].
   
\bibitem{Giunti:2007ry} 
  C.~Giunti and C.~W.~Kim,
  \emph{Fundamentals of Neutrino Physics and Astrophysics},
   Univ. Pr., Oxford 2007
\bibitem{Bourrion:2015axa} 
  O.~Bourrion {\it et al.},
  \emph{Trigger and readout electronics for the STEREO experiment},
  JINST {\bf 11}, C02078 (2016)
  [arXiv:1510.08238].
\bibitem{Buck:2015clx} 
  C.~Buck, A.~P.~Collin, J.~Haser and M.~Lindner,
  \emph{Investigating the Spectral Anomaly with Different Reactor Antineutrino Experiments},
  [arXiv:1512.06656].  
\end{thebibliography}
\end{document}